\documentclass[osajnl,twocolumn,showpacs,superscriptaddress,10pt]{revtex4-1} 

\usepackage{amsmath}    
\usepackage{graphicx}   
\usepackage{color}
\usepackage[10pt]{moresize}
\usepackage{amsfonts,amssymb,amscd,amsmath}
\usepackage{enumerate}
\usepackage{epsfig}
\usepackage{subfigure}
\usepackage{graphicx}
\usepackage{bm}

\newcommand{\ket}[1]{\mbox{$\left| #1 \right\rangle$}}

\begin{document}

\title{Quantum key distribution with key extracted from basis information}


\author{Xiongfeng Ma}
\email{Corresponding author: xma@tsinghua.edu.cn}
\address{Center for Quantum Information, Institute for Interdisciplinary Information Sciences, Tsinghua University, Beijing 100084, China}

\begin{abstract}
In conventional quantum key distribution protocols, the secure key is normally extracted from the measurement outcomes of the system. Here, a different approach is proposed, where the secure key is extracted from the measurement bases, rather than outcomes. Compared to the original Bennett-Brassard-1984 protocol, the proposed protocol involves no hardware change but modifications in data postprocessing.  We show that this protocol is more robust against detector efficiency attacks and photon-number-splitting attacks when practical detectors and photon sources are used.
\end{abstract}


\maketitle 

Since early civilizations, every advance in encryption has been defeated by advances in hacking, often with severe consequences. Quantum cryptography \cite{Bennett:BB84:1984,PhysRevLett.67.661} holds the promise to end the hacking-defending battle by offering unconditional security \cite{Mayers:Security:2001,Lo26031999,PhysRevLett.85.441} when ideal single-photon sources and detectors are employed. In practice, ideal devices never exist, and detection loopholes and source imperfections have become the targets of various attacks \cite{Brassard:PNS:2000,Qi:TimeShift:2007,Lydersen:Hacking:2010}.

The detection efficiency loophole is first discovered in the context of nonlocality tests, such as Bell's inequality test \cite{Bell:Ineq:1964}, which are designed to disprove the theory of local hidden variables. The loophole allows local hidden variables to reproduce the prediction of the quantum theory when the detector efficiency is low enough \cite{PhysRevD.2.1418}. Nonlocality test leads to an important concept in quantum mechanics --- entanglement, which has been shown to be the precondition for the security of a quantum key distribution (QKD) system  \cite{PhysRevLett.92.217903}. Intuitively, one would expect the efficiency loophole may introduce security issues for QKD. In fact, such suspicion has been proven true by quantum hacking \cite{Qi:TimeShift:2007}. Various approaches have been proposed to close this loophole \cite{PhysRevLett.97.120405,MXF:DDIQKD:2012}.

One way to close the loopholes introduced by the device imperfections is to implement the fully device-independent quantum key distribution (DIQKD) protocol \cite{mayers1998quantum,PhysRevLett.97.120405}. However, such protocol is a challenge for practical implementation due to its high requirement on physics devices. So far, no DIQKD experiment has been demonstrated even under a lab condition. Recently, Lo, Curty and Qi proposed a novel scheme, measurement-device-independent quantum key distribution (MDIQKD), to close all loopholes existing in the measurement devices \cite{PhysRevLett.108.130503}. However, it requires coincident detections and interference of two independent photon sources, which make the experiment realization challenging \cite{PhysRevLett.111.130501,PhysRevLett.111.130502,PhysRevLett.112.190503}. Also, the security of the MDIQKD protocol relies on trustful implementation of source encoding.

On the source side, the loophole exists in imperfect single photon sources, which inevitably emit multi photon states. From the study of photon-number-splitting (PNS) attacks \cite{Huttner:PNS:1995,Brassard:PNS:2000}, multi photon states would cause severe security issues for QKD. Such problem can be solved by introducing decoy states in the system \cite{Hwang:Decoy:2003,Lo:Decoy:2005,Wang:Decoy:2005}. Note that the decoy-state method is applied to MDIQKD protocol \cite{PhysRevLett.108.130503}.

In this work, we propose a QKD protocol that is not only able to close the detection efficiency loophole, but also make the system more robust against imperfect source attacks. Compared to the current QKD realizations, the protocol involves no hardware change but modifications in data postprocessing. Thus, it offers immediate applications in quantum cryptography.

The protocol, between two legitimate users Alice and Bob, runs as follows.
\begin{enumerate}
\item
Alice randomly prepares one of the four BB84 states, $\ket{0}_x$, $\ket{1}_x$, $\ket{0}_z$, or $\ket{1}_z$, and sends the state to Bob.

\item
Upon receiving the state, Bob randomly chooses the $X$ or $Z$ basis for measurement.

\item
Alice and Bob record their raw key bits according to the basis they have chosen. Alice sets her key bit to be 0 for the two states in the $Z$ basis, $\ket{0}_z$ and $\ket{1}_z$, and 1 for the two states in the $X$ basis, $\ket{0}_x$ and $\ket{1}_x$. Bob sets his key bit to be 0 for the $X$ basis measurement and 1 for the $Z$ basis measurement.

\item
Bob publicly announces the measurement outcomes, $0$ (for $\ket{0}_x$ or $\ket{0}_z$) or $1$ (for $\ket{1}_x$ or $\ket{1}_z$), \emph{but keeps his basis information confidential}.

\item
Alice and Bob performs raw key sift: they discard the key when Bob's measurement outcome matches Alice's state. For example, if Alice sends out a state, $\ket{0}_x$ or $\ket{0}_z$, and Bob announces 0, they will discard the raw key bit.

\item
They perform error correction and privacy amplification on the sifted key bits to extract a final secure key.
\end{enumerate}

One can see that the first two steps form the quantum phase of the proposed protocol, which is exactly the same as the regular BB84 protocol. Thus, there is no hardware change required to implement the new protocol. Note that in Step 4, the measurement outcome announcement can also be done by Alice. The definition of bit 0 and 1 are different by Alice and Bob. It is not hard to see that the raw key sift factor is 1/4, comparing to 1/2 in the case of the original BB84.

Let us first compare the underlying assumptions in the MDIQKD protocol and the new protocol. In the MDIQKD protocol, the whole measurement device is assumed to be in the hand of an adversary, Eve, who might not actually perform the measurement as designed or honestly announce the true outcomes. Here in the new protocol, Bob still trusts the measurement device such as the basis control. Eve can manipulate the measurement results by controlling the detector efficiencies, such as the time-shift attack \cite{Qi:TimeShift:2007}, but she can only control the detectors in a basis independent manner. That is, the detectors respond the same to the two bases. Since the final measurement outcomes are publicly announced, Eve can also learn the detection results, but not the basis information. The new protocol still suffers from the basis-dependent attacks, such as the strong pulse attack \cite{Lutkenhaus:DoubleClick:1999} and the strong illumination attack \cite{Lydersen:Hacking:2010}. From this point of view, the MDIQKD protocol requires less assumption on detection devices than the new protocol.

On the source side, due to the PNS attacks \cite{Huttner:PNS:1995,Brassard:PNS:2000}, the 2-photon state is not secure for the MDIQKD protocol. Later, we will show that this is not the case for the new protocol. Thus, the new protocol is more robust against source attacks than the MDIQKD protocol.

\begin{figure}[hbt]
\centering \resizebox{8cm}{!}{\includegraphics{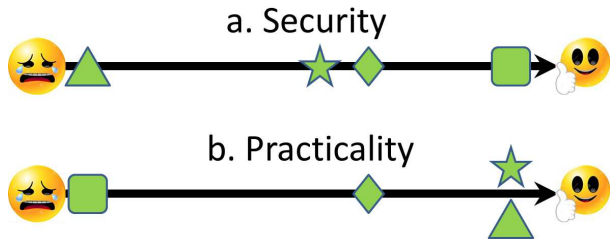}}
\caption{Comparison of four QKD protocols in two aspects: security and practicality. $\blacktriangle$: current QKD realization; $\bigstar$: the new QKD protocol; $\blacklozenge$: MDIQKD; $\blacksquare$: DIQKD.} \label{Fig:ProsCons}
\end{figure}

From the practical point of view, a MDIQKD system requires coincident detection, while the hardware part of the new protocol is the same as the regular BB84 protocol. Thus, the new protocol is more practical than the MDIQKD protocol. We can put all these pros and cons for the new protocol and the MDIQKD protocol, along with the current QKD realizations and the fully device-independent one (DIQKD)\cite{PhysRevLett.97.120405}, in Fig.~\ref{Fig:ProsCons}, from where one can see that the new protocol enjoy the both side of the worlds: security and practicality.

Now, let us take a look at how the proposed protocol is secure against the known attacks. We will leave the full security proof in future works.

In an intercept-and-resend attack, Eve measures the state randomly in the $X$ or $Z$ basis and sends qubits to Bob according her measurement outcomes. Without loss of generality, we assume Alice sends out state $\ket{0}_z$ in the $Z$ basis, corresponding to a key bit value of 0. The result is shown in Table \ref{Tab:DDI:IaR}. One can see that like BB84, Eve will not introduce any disturbance when she chooses the same basis as Alice, but she has 50\% chance to introduce an error when she picks up the wrong basis. Thus, the new protocol is secure against the simple intercept-and-resend attack.

\begin{table}
\centering
\caption{Alice sends out state $\ket{0}_z$ (corresponding to a key bit 0). Eve performs a simple intercept-and-resend attack. From left to right, Eve randomly chooses a measurement basis $M_{Eve}=X$ or $Z$. Then she resends her measurement result to Bob. Bob randomly chooses two bases, $X$ (bit 0) or $Z$ (bit 1) for measure. Obviously, when Eve chooses the same basis as Alice, she will not introduce any disturbance.} \label{Tab:DDI:IaR}
\begin{tabular}{cc|ccccccc}
\hline
\hline
$M_{Eve}$  & resend & $M_{Bob}$ & outcome & result & disturbance \\
\hline
$Z$ & $\ket{0}_z$ & - & - & - & none \\
$X$ & $\ket{0}_x$ & X & $\ket{0}_x$ & inconclusive & none \\
    &  & Z & $\ket{0}_z$ & inconclusive & none \\
    &  & Z & $\ket{1}_z$ & 1 & error! \\
    & $\ket{1}_x$ & X & $\ket{1}_x$ & 0 & none \\
    &  & Z & $\ket{0}_z$ & inconclusive & none \\
    &  & Z & $\ket{1}_z$ & 1 & error! \\
\hline
\hline
\end{tabular}
\end{table}

By going through the rest three cases, where Alice sends $\ket{1}_z$, $\ket{0}_x$, and $\ket{1}_x$, one can see that if Eve chooses the same basis as Alice or Bob, she would not introduce any disturbance. If Alice and Bob choose the same bases and Eve chooses the other basis, she would introduce an error when the result is conclusive.

As shown in Table \ref{Tab:DDI:IaR}, when Eve chooses the same basis as Alice, the raw key sift factor is 1/4 as discussed previously. When Eve chooses the different basis as Alice, the raw key sift factor becomes 1/2, in which case there is 50\% error rate. Thus, the total error rate introduced by Eve if she performs the simple intercept-and-resend attack will be 1/3, which is different from 1/4 as the regular BB84 protocol.

We assume Bob uses two detectors to distinguish two states, $\ket{0}$ and $\ket{1}$, after the measurement basis is set. In the efficiency control attack, we assume Eve has a full control of the detector efficiency after the basis is set (basis independent).  That is, she can make detectors for $\ket{0}$ and $\ket{1}$ active (100\% efficiency) and inactive (0\% efficiency). This is the extreme case of the time-shift attack \cite{Qi:TimeShift:2007}. Since the detection results are announced publicly and also the detector respond the same to different basis choices, such attack is ineffective.

In the PNS attack, Eve splits the 2-photon state, and stores one photon in her memory and sends the rest to Bob. After Alice and Bob compare the measurement outcomes, Eve would measure her photon. If Eve chooses the same basis as Alice, she would not get a conclusive result. If she chooses the same basis as Bob, she has a 50\% chance to get a conclusive result. Thus, Eve's information about the key is not full. Note that for higher number Fock states, such as 3-photon state, the protocol is not secure due to the unambiguous state discrimination (USD) attack \cite{Lutkenhuas:USD:2000}. This is similar to the SARG04 protocol \cite{PhysRevLett.92.057901}. The key difference is that the SARG04 protocol extract key information from the measurement outcomes, which makes it suffers from the detector efficiency mismatch attacks. While in the proposed protocol, the key is extracted from the measurement basis, which makes it robust against detector efficiency mismatch attacks.

In practice, due to detector efficiency mismatch or detector dead time, the raw data from a QKD system normally cannot pass the statistical randomness tests. With the proposed protocol, since its key information comes from the measurement basis choice, the raw data can easily pass the randomness tests.


This protocol can be proven to be secure, following the security proof~\cite{Tamaki:2006:SARG04,*Fung:2006:SARG04} of the SARG04 protocol which uses non-orthogonal states to represent bits ``0'' and ``1''. In fact, the protocol proposed here may also be regarded as such, using two sets of non-orthogonal states instead of four in SARG04. To prove the security, we form an equivalent entanglement-distillation protocol for the current QKD protocol by capturing the QKD operations in a joint state of Alice and Bob. To do this neatly, we construct a filtering operation to be performed by Bob. This operation converts non-orthogonal states to orthogonal ones and can be shown to be equivalent to Bob's USD measurement. We can then use this joint state of Alice and Bob to determine the desired security relation, which is obtained by projecting this joint state onto the Bell basis. The outcomes give a relation between the bit and phase error probabilities (denoted as $e_b$ and $e_p$ respectively). In the experiment, we can estimate the bit error rate directly, and from this relation we can upper bound the phase error rate as well. Finally, we can choose a quantum error correcting code tolerable to these two error rates following Shor-Preskill's proof~\cite{PhysRevLett.85.441} and derive a secret key. The related work is in progress.

In summary, we have presented a QKD protocol that is robust against detection efficiency loophole and PNS attacks. Compare to the current BB84 realizations, the proposed protocol do not involve any hardware changes. This scheme can be directly applied to current practical QKD systems, with modifications in the data postprocessing procedure. The security of the protocol lies on the assumption that the detectors behave independently of Bob's basis choice. Combining with the decoy-state method, the proposed method would offer a practical solution to the secure information exchange in practice. We remark that this technique used in the proposed QKD protocol may also be applicable to non-locality tests.

After completing this manuscript, we notice a few related works are put on online pre-print \cite{gonzalez2014quantum,lim2014detector,cao2014highly,qi2014trustworthiness}.

The author acknowledges insightful discussions with C.-H.~F.~Fung, H.-K.~Lo, and X.~Yuan. This work was supported by the National Basic Research Program of China Grants No.~2011CBA00300 and No.~2011CBA00301, and the 1000 Youth Fellowship program in China.

\bibliographystyle{apsrev4-1}

\bibliography{BibDDIQKD}

\end{document}